\documentclass[a4paper,11pt]{article}
\pdfoutput=1
\usepackage{jcappub}
\usepackage[T1]{fontenc} 
\usepackage{hyperref}
\usepackage{graphics}
\usepackage{amssymb, amsmath}
\usepackage[dvipsnames]{xcolor}
\usepackage{hhline}
\usepackage{multirow}
\usepackage{placeins}
\usepackage{afterpage}
\usepackage{xcolor,color,colortbl}

\author[a,b,c,d]{Alex Krolewski}
\emailAdd{akrolewski@perimeterinstitute.ca}
\author[c,d]{Simone Ferraro}
\emailAdd{sferraro@lbl.gov}

\affiliation[a]{AMTD Fellow, Waterloo Centre for Astrophysics, University of Waterloo, Waterloo ON N2L 3G1, Canada}
\affiliation[b]{Perimeter Institute for Theoretical Physics, 31 Caroline St. North, Waterloo, ON NL2 2Y5, Canada}
\affiliation[c]{Physics Division, Lawrence Berkeley National Laboratory, Berkeley, CA 94720}
\affiliation[d]{Berkeley Center for Cosmological Physics, University of California, Berkeley, CA 94720}

\title{The Integrated Sachs Wolfe effect: unWISE and Planck constraints on Dynamical Dark Energy}

\keywords{CMB}

\abstract{CMB photons redshift and blueshift
as they move through gravitational potentials $\Phi$
while propagating across the Universe.  If the potential is not constant in time, the photons will pick up a net redshift or blueshift, known as the Integrated Sachs-Wolfe (ISW) effect.
In the $z \ll 1000$ universe, $\dot{\Phi}$ is nonzero on large scales when the Universe transitions from matter to dark energy domination. This effect is only detectable in cross-correlation with large-scale structure at $z \sim 1$. In this paper we present a 3.2$\sigma$ detection of the ISW effect using cross-correlations between unWISE infrared galaxies and Planck CMB temperature maps. We use 3 tomographic galaxy samples spanning $0 < z < 2$, allowing us to fully probe the dark energy domination era and the transition into matter domination. This measurement is consistent with $\Lambda$CDM ($A_{\rm ISW} = 0.96 \pm 0.30$).
We study constraints on a particular class of dynamical dark energy models
(where the dark energy equation of state is different in matter and dark energy domination),
finding that unWISE-ISW improves constraints
from type Ia supernovae due to improved constraints on the time evolution of dark energy. When combining with BAO measurements, we obtain the tightest constraints on specific dynamical dark energy models.
In the context of a phenomenological model for freezing quintessence, the Mocker model, we constrain the dark energy density within 10\% at $z < 2$ using ISW, BAO and supernovae. Moreover, the ISW measurement itself provides an important independent check when relaxing assumptions about the theory of gravity, as it is sensitive to the gravitational potential rather than the expansion history.}

\arxivnumber{20MM.NNNNN}

\begin{document}
\maketitle
\flushbottom

\section{Introduction}

On super-horizon scales ($\ell < 200$) the cosmic microwave
background (CMB) power spectrum is dominated by the Sachs-Wolfe effect \citep{SachsWolfe67,ReesSciama68}.
On these scales, fluctuations in the gravitational potential generate fluctuations
in the CMB as photons exiting the potentials are redshifted and blueshifted.
For adiabatic fluctuations in a matter-dominated universe, the temperature fluctuations
generated by the Sachs-Wolfe effect are
\begin{equation}
\frac{\Delta T}{T} = - \frac{1}{3} \Phi
\label{eqn:deltaT}
\end{equation}
where the factor of $-1/3$ comes from gravitational redshifting as photons
exit the potential (contributing $-\Phi$) and clocks running slow within the gravitational
potential (contributing $2\Phi/3 $) \citep{SachsWolfe67,WhiteHu96}.
Since the dimensionless angular power spectrum of potential fluctuations is flat, this creates a characteristic ``Sachs-Wolfe plateau''
at low $\ell$ in the primary CMB.

The Integrated Sachs-Wolfe effect is an additional effect from the time dependence of the gravitational potential as photons propagate through the observable universe, from the surface of last scattering to the observer on Earth.
In matter domination and in the linear regime, gravitational potentials are constant in time, so photons redshift in and out of potentials with no effect. However, the gravitational
potential decays during dark energy domination, and this leads to a net blueshift of photons as they travel through decaying potentials.
This effect is too small to be detectable in the CMB power spectrum, but is detectable in cross-correlation with large-scale structure. 

Since the ISW effect in linear theory is only sourced by dark energy, it is a powerful direct probe of it, complementary to measurements based on the distance-redshift
relation, matter clustering, and the primary CMB \citep{Riess98,Perlmutter99,Riess04}.  Moreover, while often measurements are ``integrated'' over a large redshift range, the dependence of ISW on $\dot{\Phi}$ at the redshift of the measurement only can be more sensitive to variations in dark energy properties over relatively narrow redshift ranges.
As a result, measurements of the ISW effect can test dark energy \citep{CrittendenTurok96} and probe extensions to $\Lambda$CDM
such as modified gravity \citep{Hu02,Renk17} or spatial curvature \citep{Kamionkowski96}.

On nonlinear scales, even during matter
domination, gravitational potentials are not
constant in time. This additional nonlinear
integrated Sachs-Wolfe effect is also known
as the Rees-Sciama effect \citep{ReesSciama68,Cooray02} but it is generally
too small to detect with the current generation of surveys, and is found at larger $\ell$ than
considered here. In addition, a time-varying potential is also generated by the galaxy motion across the line of sight \cite{1983Natur.302..315B}, an effect known as ``moving lens'' or ``slingshot''. This component of the signal and its detectability has been studied in \cite{Stebbins:2006tv, Hotinli:2018yyc, Hotinli:2020ntd, Hotinli:2021hih, Yasini:2018rrl, Hagala:2019ncx}, and we note that a simple cross-correlation between galaxy positions and the CMB fluctuations will not receive a contribution from the ``moving lens'' effect, and therefore we won't consider it here.

Since the first detection of the ISW cross-correlation by Ref.~\cite{BoughnCrittenden04},
there have been many ISW detections with a variety of galaxy samples and CMB data from WMAP \citep{Scranton03,Fosalba03,Nolta04,Corasaniti05,Padmanabhan05,
Vielva06,Giannantonio06,Cabre07,Rassat07,McEwen07, Ho08, Giannantonio12,Goto12,Hernandez14,Ferraro15,MouraSantos16,Shajib16} and Planck \citep{Planck2013_ISW,Planck2015_ISW,Stolzner18,Ansarinejad20,Hang20}.
The highest significance detections range from 4--5$\sigma$ \citep{Giannantonio08,Planck2015_ISW,Stolzner18}.
These measurements have been used to constrain cosmological parameters and dark energy, including the curvature of the universe
and 
the dark energy equation of state \citep{Nolta04,Pietrobon06,McEwen07,Vielva06,Giannantonio08,Ho08,Xia09,Zhao10,LiXia10}. 

The unWISE catalog \citep{Schlafly19} is ideal for an ISW cross-correlation measurement. It contains $\sim 500$ million galaxies across the full
sky out to $z \sim 2$, covering the entire dark-energy dominated epoch. We split the sample into three redshift bins using the unWISE galaxies' infrared colors, allowing us to probe the ISW tomographically.

We point out that the area and redshift coverage of this work is significantly increased over previous analyses, which are either close to full-sky but at lower redshift (i.e.\ 2MASS \cite{Rassat07}, WISE-Galaxy \cite{Ferraro15,Shajib16,Planck2015_ISW}) or with broad redshift kernels (WISE-AGN \cite{Ferraro15, Shajib16, Planck2015_ISW}, NVSS \cite{Giannantonio08}). The sky and redshift coverage of the lower redshift bins are ideal for overlap with the ISW kernel. The highest redshift bin adds little signal in the standard $\Lambda$CDM model because it is mostly in the matter-dominated era.  However, even in absence of a strong detection, it will provide very important information about possible deviations of dark energy from a cosmological constant during the transition between matter and dark energy domination, typical of the models that we consider in this work.

Type Ia supernovae and baryon acoustic
oscillations (BAO) generally constrain dark energy
at $z < 1$, leaving the dark energy
equation of state at higher redshift
relatively unconstrained.  Tomography over the range $0 \lesssim z \lesssim 2$ is particularly interesting in light of the reported ISW ``anomalies'' obtained when stacking on supersclusters and supervoids, where an anomalous signal is found \citep{Grannett08,Kovacs19, Kovacs21}, and the size (as well as the sign) of this discrepancy evolves quite rapidly over this redshift range.  For example, there have been hints of a negative ISW signal at $z \sim 1.5$ \cite{Kovacs21}, which should
leave an imprint in our measurement out to $z = 2$.

To demonstrate how this ISW measurement
improves our knowledge of dark energy at $z > 1$, we consider constraints on dynamical dark energy models, i.e.\ models where the
dark energy equation of state changes at $z \sim 1-2$ (see for example \cite{Bull21, Linder06, Linder:2006ud}).
These are sometimes referred to in the literature as
``early dark energy'' models but should not be confused
with models postulating an additional dark energy component in the $z \sim 1000$ universe in order to address the Hubble tension \cite{Poulin19,Agrawal19,Lin19,Smith20,Hill21}.\footnote{However, note that some of the constraining power from the CMB on these
$z \sim 1000$ early dark energy models
comes from the early ISW effect: the decay of potentials
at the transition from radiation to matter domination
\cite{Vagnozzi21,Hill21}.}
To avoid confusion, we will refer to the models
with dark energy decaying at $z \sim 1-2$ as ``dynamical
dark energy'' throughout this paper.

The outline of the paper is as follows.
We describe the theory
in Section~\ref{sec:isw_theory}, summarize
the data used in Section~\ref{sec:isw_data},
describe the methodology of the ISW cross-correlation
measurement in Section~\ref{sec:isw_measurements},
and present the measurement and compare to $\Lambda$CDM and a freezing quintessence model for dynamical dark energy in Section~\ref{sec:ISW_measurement}.
Where necessary we assume a fiducial $\Lambda$CDM cosmology with the Planck 2018 maximum likelihood parameters, $\Omega_m = 0.3096$, $H_0 = 67.66$,
$n_s = 0.9665$, $\sigma_8 = 0.8102$, $\Omega_b = 0.049$, and one massive neutrino with mass 0.06 eV.
We quote magnitudes in the Vega system, noting that we can easily convert these to AB magnitudes with AB = Vega + 2.699, 3.339 in W1, W2, respectively.

\section{Theory}
\label{sec:isw_theory}

The integrated Sachs-Wolfe effect comes from the blueshifting of CMB photons due to a changing gravitational potential
\begin{equation}
\left(\frac{\Delta T}{T} \right) _{\rm ISW} = - 2 \int d\chi \ e^{-\tau} \dot{\Phi}
\label{eqn:isw_temp}
\end{equation}
where the factor of 2 comes from the fact that both the spatial and time components of the perturbed potential
contribute to the ISW effect, $\chi$ is the comoving distance, $\tau$ is the optical depth to distance $\chi$ and the dot refers to derivative with respect to conformal time. Since we will only work with low redshift samples, we can neglect the $e^{-\tau}$ term and set it to 1.  In the linear regime and after recombination, $\dot{\Phi}$ is only nonzero in dark energy domination.

Since the ISW is the only physical correlation between foreground galaxies and CMB temperature at low $\ell$,
the cross-power spectrum $C_{\ell}^{Tg}$ is given by
\begin{equation}
C_{\ell}^{Tg} = C_{\ell}^{\dot{\Phi} g} = \frac{2}{\pi} \int \, k^2 \, dk \, P(k) K_{\ell}^{\dot{\Phi}}(k) \, K_{\ell}^g(k)
\label{eqn:isw}
\end{equation}
The kernel functions $K_{\ell}^{\dot{\Phi}}(k) $ and $K_{\ell}^g(k)$ are
\begin{equation}
K_{\ell}^g(k) = \int \, dz \, b(z) \frac{dN}{dz} D(z) \, j_{\ell}[k \chi(z)]
\label{eqn:gal_kernel}
\end{equation}
\begin{equation}
K_{\ell}^{\dot{\Phi}}(k) = \frac{3 \Omega_m H_0^2}{k^2} \int dz\, \frac{d}{dz}\left[ (1+z) D(z) \right] \, j_{\ell}[k \chi(z)]
\label{eqn:isw_kernel}
\end{equation}
where $dN/dz$ is the galaxy redshift distribution, $b(z)$ is the bias evolution, $D(z)$ is the linear growth factor,
and $j_{\ell}$ are spherical Bessel functions.
We compare the ISW sensitivity, $\frac{d[(1+z) D(z)]}{dz} \frac{dV}{dz}$, to the redshift distribution
of the unWISE galaxies in Fig.~\ref{fig:dndz}.

\begin{figure}
    \centering
    \resizebox{1.0\columnwidth}{!}{\includegraphics{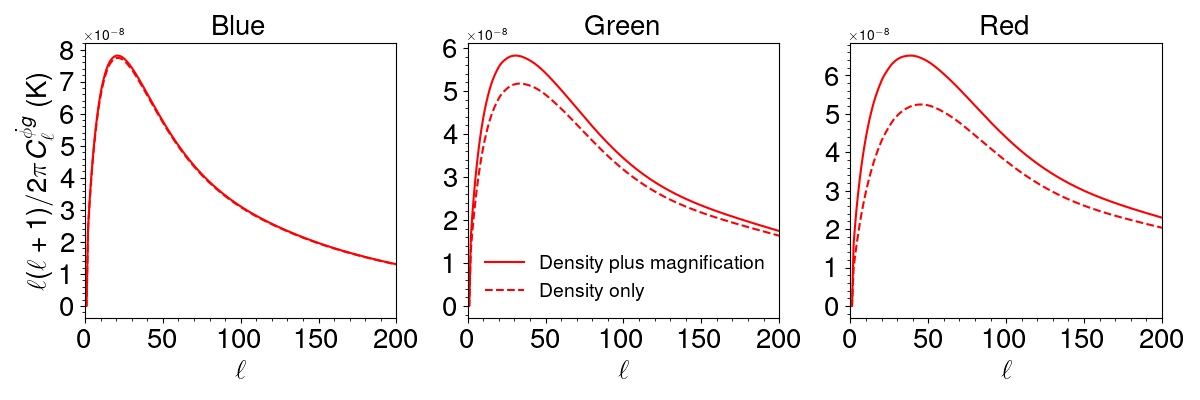}}
    \caption{ISW-unWISE cross-correlation in the fiducial flat $\Lambda$CDM cosmology.  The contributions
    from the density, $C_{\ell}^{\dot{\phi} g}$,
    and magnification $C_{\ell}^{\dot{\phi} \mu}$
    are plotted separately.  All of the other higher-order terms (i.e.\ redshift space distortions, peculiar velocity, time delay,
    source evolution, ISW, time delay; see ref.~\cite{Challinor:2011bk} for a complete discussion) change the total by $< 2\% $ ($< 0.3\%$ at $\ell > 20$), and are largest for blue and red.
    }
    \label{fig:isw_theory_curves}
\end{figure}

We also consider the cross-correlation between the ISW and cosmic magnification, $C_{\ell}^{\dot{\Phi} \mu}$
\begin{equation}
\label{eqn:isw_mag}
C_{\ell}^{T\mu} = C_{\ell}^{\dot{\Phi} \mu} = \frac{2}{\pi} \int \, k^2 \, dk \, P(k) K_{\ell}^{\dot{\Phi}}(k) \, K_{\ell}^{\mu}(k)
\end{equation}
with kernel $ K_{\ell}^{\mu}(k)$
\begin{equation}
K_{\ell}^{\mu}(k) = (5s-2)\,\frac{3}{2} \Omega_m H_0^2   \int dz \,  (1+z) g_i(\chi(z)) D(z) j_{\ell}[k \chi(z)]
\end{equation}
where $s$ is the response of the number density to magnification, and 
\begin{equation}
g_i(\chi) = \int_{\chi}^{\chi_{\star}} d\chi' \ \frac{\chi(\chi' - \chi)}{\chi'} \ H(z') \ \frac{dN_i}{dz'}
\end{equation}
For the green and red samples, $s$ is large
enough that magnification bias makes a substantial
contribution to the observed unWISE-CMB temperature
cross-correlation (Fig.~\ref{fig:isw_theory_curves}).

Note that Eqs.~\ref{eqn:isw}--\ref{eqn:isw_kernel} are only valid in the linear regime where the power spectrum $P(k,z)$ can be separated into $P(k)$ and $D(z)$.
To improve accuracy at higher $\ell$, we  instead use the nonlinear $P(k)$ \citep{Cai10,Hang20} from \textsc{HALOFIT}
\citep{Mead15},
although we find this makes very little difference on the large scales we consider.
We use \textsc{CAMBSources} \cite{Challinor:2011bk}, part of the \textsc{CAMB} package \cite{Lewis:1999bs,Howlett:2012mh}, to compute $C_{\ell}^{Tg}$.

\section{Data}
\label{sec:isw_data}

\subsection{unWISE}
\label{sec:unwise}

\subsubsection{Galaxy catalog}

The WISE mission mapped the entire sky at 3.4, 4.6, 12 and 22 $\mu$m (W1, W2, W3, and W4), with angular
resolutions of 6.1'', 6.4'', 6.5'' and 12'', respectively \citep{Wright10}.  The original mission collected data
in 2010. After a two year hibernation, the satellite
was reactivated and continued observations as NEOWISER (NEOWISE-Reactivation) \citep{Mainzer11,Mainzer14} and has been continuously mapping
the sky in W1 and W2 since 2014.  The unWISE catalog
was created from the first five years of imaging: one year of WISE and four from NEOWISE \citep{fulldepth_neo1,fulldepth_neo2,fulldepth_neo3,Meisner19}. It reaches 0.7 mags deeper than the AllWISE
catalog from 2010.

The W1 and W2 magnitudes enable
us to roughly divide the sample by redshift, yielding 3 samples spanning $0 < z < 2$. Additionally, we use
Gaia to remove stars from the sample, yielding residual
stellar contamination of $\sim 2\%$ as measured
by deep imaging in the COSMOS field \citep{Laigle+15}.
These samples were characterized and used in Ref.~\cite{Schlafly19,Krolewski20,Krolewski21}, and we refer the reader to Ref.~\cite{Krolewski20} for a more comprehensive discussion.
We reproduce
Table \ref{tab:galaxyselection_isw} from Refs.~\cite{Krolewski20,Krolewski21} to summarize the important properties of the sample: the color selection, redshift distribution,
number density, galaxy bias, and 
response of number density to galaxy magnification $s \equiv d\log_{10}N/dm$. We measure $s$ using galaxies with ecliptic latitude $|\lambda| >60^{\circ}$, where the WISE depth of coverage
is greater and thus the measurement of $s$ is less affected
by incompleteness (see discussion in Appendix D
in Ref.~\cite{Krolewski20}).

\begin{table}[]
\small
\centering
\begin{tabular}{c|ccccccccc}
Label & $\mathrm{W1}-\mathrm{W2} > x$ & $\mathrm{W1}-\mathrm{W2} < x$ & $\mathrm{W2} < x$ & $\bar{z}$ & $\delta z$ & $\bar{n}$  & $s$ & $b^{\rm eff} \pm \textrm{stat} \pm \textrm{$dN/dz$}$ \\
\hline
Blue & & $(17-\mathrm{W2})/4+0.3$ & 16.7 & 0.6 & 0.3 & 3409 & 0.455 & $1.50 \pm 0.025 \pm 0.037$ \\
Green & $(17-\mathrm{W2})/4 + 0.3$ & $(17-\mathrm{W2})/4 + 0.8$ & 16.7 & 1.1 & 0.4 & 1846  & 0.648 & $2.23 \pm 0.032 \pm 0.025$ \\
Red & $(17 - \mathrm{W2})/4 + 0.8$ & & 16.2 & 1.5 & 0.4 & 144  & 0.842 & $3.19 \pm 0.076 \pm 0.059$ \\
\end{tabular}
\caption[Summary of unWISE galaxy samples]{WISE Color and magnitude cuts for selecting unWISE galaxies of different redshifts, together with the mean redshift, $\bar{z}$, and the width of the redshift distribution, $\delta z$ (as measured by matching to objects with photometric redshifts on the COSMOS field \cite{laigle:2016}); number density per deg${}^2$ within the unWISE mask, $\bar{n}$; response of the number density to magnification, $s \equiv d\log_{10}N/dm$;
and average bias $b^{\rm eff}$ (Eq.~\ref{eqn:beff}) along with its statistical and systematic errors from uncertainty in the redshift distribution.
Galaxies are additionally required to have $\mathrm{W2} > 15.5$, to be undetected or not pointlike in Gaia, and to not be flagged as diffraction spikes, latents or ghosts. See Refs.~\cite{Schlafly19,Krolewski20,Krolewski21} for further details.
}
\label{tab:galaxyselection_isw}
\end{table}

We require that the blue and green samples have $15.5 < \mathrm{W2} < 16.7$, and the red sample has $15.5 < \mathrm{W2} < 16.2$;
in Ref.~\cite{Krolewski20} we find that deeper red samples are potentially affected by systematics.

We remove potentially spurious sources (diffraction spikes, latents, ghosts) and 
all galaxies are required to be either undetected or not pointlike in Gaia.  Here a source is taken as ``pointlike'' if 
\begin{equation}
\mathrm{pointlike}(G, A) = \begin{cases}
\log_{10} A < 0.5 & \text{if $G < 19.25$} \\
\log_{10} A < 0.5 + \frac{5}{16} (G-19.25) & \text{otherwise} \, ,
\end{cases}
\end{equation}
where G is the Gaia G band magnitude and $A$ is \texttt{astrometric\_excess\_noise} from Gaia DR2 \cite{Gaia18}.  A source is considered ``undetected'' in Gaia if there is no Gaia DR2 source within $2.75''$ of the location of the WISE source.  High \texttt{astrometric\_excess\_noise} indicates that the Gaia astrometry of a source was more uncertain than typical for resolved sources; this cut essentially takes advantage of the $0.1''$ angular resolution of Gaia to morphologically separate point sources from galaxies.  We estimate that this reduces the stellar contamination in our samples to $< 2\%$.

\begin{figure}
    \centering
    \hspace{-40pt}
    \resizebox{1.1\columnwidth}{!}{\includegraphics{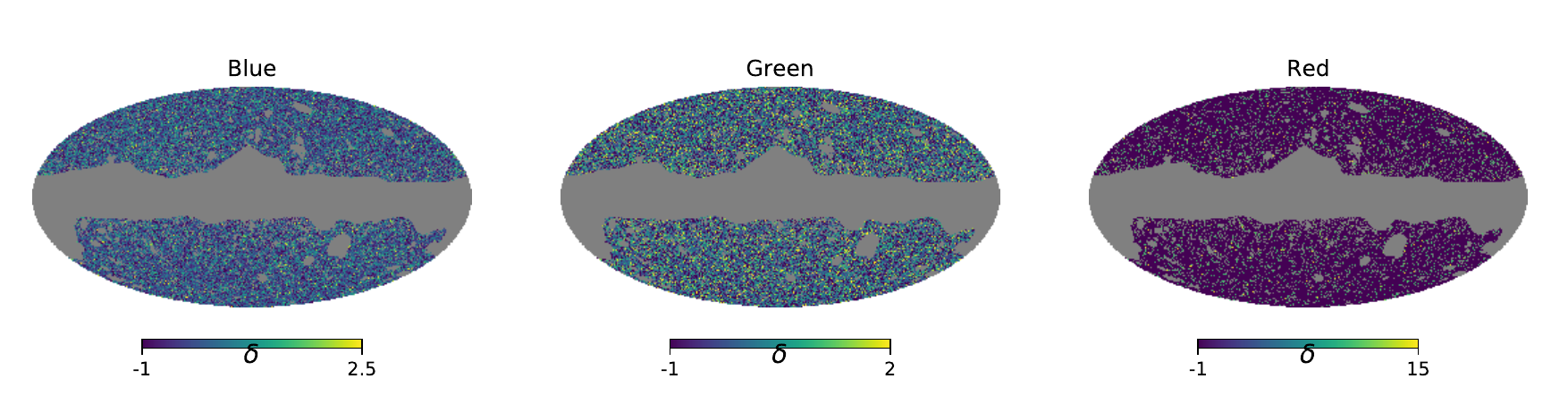}}
    \caption{Sky distribution of the 3 unWISE samples,
    in Galactic coordinates.
    }
    \label{fig:unwise_maps}
\end{figure}

The unWISE mask is based on the 2018 Planck lensing mask \citep{PlanckLens18}.  We additionally mask a small portion of the sky at $|b| < 10^{\circ}$,
and mask bright infrared stars, diffraction spikes, nearby galaxies, planetary nebulae, and low latitude pixels with a substantial number of fainter stars which
will reduce the effective area in a pixel by masking galaxies within 2.75'' of each star (due to the Gaia criterion).  The full details of the mask construction are in section 2.3 of
Ref.~\cite{Krolewski20}; this mask yields $f_{\rm sky} = 0.586$. Sky distributions of the masked unWISE
galaxy samples are shown in Fig.~\ref{fig:unwise_maps}.

We also create systematics
weights for the unWISE galaxy
samples to remove correlations
with potentially contaminating
large-scale systematics, such
as Milky Way stellar density
or WISE depth-of-coverage.
We follow a similar methodology
to the linear regression
method used for SDSS and DES
galaxy clustering analysis
\citep{Ross12,Ross17,Ata18,Bautista18,Elvin-Poole18,Ross20,RodriguezMonroy21}.
We start by creating HEALPix maps \cite{Gorski05} at NSIDE=128 of the unWISE galaxy density field;
stars from the Gaia catalog;
unWISE W1 and W2 5$\sigma$ limiting magnitude;
dust extinction $E(B-V)$ from the Schlegel-Finkbeiner-Davis
map \citep{SFD};
neutral hydrogen column density $N_{\textsc{HI}}$
from the H14PI survey \citep{H14PI};
and
3.5 and 4.9 $\mu$m sky brightness from the DIRBE Zodi-Subtracted Mission Average (ZSMA)\footnote{\url{https://lambda.gsfc.nasa.gov/product/cobe/dirbe_zsma_data_get.cfm}}, and a separate
estimate of the zodiacal background light from the DIRBE Sky and Zodi Atlas (DSZA)\footnote{\url{https://lambda.gsfc.nasa.gov/product/cobe/dirbe_dsza_data_get.cfm}} \citep{Kelsall98}.

For the blue sample, we fit a linear trend between unWISE galaxy density and Gaia stellar density,
and a piecewise linear trend to W2 5$\sigma$ limiting magnitude.
We determine errorbars on each binned systematic property using the variance
of density values from 100 Gaussian mocks with no correlation
with the systematics templates.
We find that correlations with all other templates (and residual
correlations with stars and W2 depth) are $\lesssim 1\%$ after weighting.
For green, the picture is similar, though we need to use a piecewise
linear fit for both stellar density and W2 5$\sigma$ magnitude.
Finally, for red we find that the most significant trends
are to stellar density and $N_{\textsc{HI}}$, and fit piecewise linear trends
to only these templates.

\subsubsection{Bias and redshift distribution}

Theory predictions for the ISW cross-correlation require both the redshift distribution of the galaxy sample, $dN/dz$, and its bias evolution $b(z)$ (Section~\ref{sec:isw_theory}).
In the equations below, we assume that $dN/dz$ has already been normalized so that $\int dz \, \frac{dN}{dz} = 1$.
Our best-characterized measurement of the unWISE redshift distribution
comes from cross-correlations with spectroscopic
galaxies and quasars from SDSS \citep{Krolewski20}.  The cross-correlation between unWISE and known
spectroscopic galaxies in a narrow redshift bin at $z_i$
is proportional to the fraction of unWISE galaxies
in that redshift bin, $dN/dz(z_i)$, and the bias
of the unWISE galaxies $z_i$, given that we can precisely
measure the bias of the spectroscopic sample
from its autocorrelation \citep{Newman08,McQuinn13,Menard13}.
Repeating this using samples spanning the unWISE
redshift range (e.g.\ galaxies and quasars from SDSS \cite{Alam20}) allows us to determine the bias-weighted redshift distribution $b(z) dN/dz$.

We normalize the cross-correlation redshift distribution
to integrate to unity,
and refer to the resulting quantity as $f(z) dN/dz$:
\begin{equation}
f(z) \frac{dN}{dz} \equiv \frac{b(z) \frac{dN}{dz}}{\int \, dz \, b(z) \frac{dN}{dz}}
\end{equation}
Measurements of the product $f(z) dN/dz$ are convenient, because they appear together in the galaxy
kernel in equation~\ref{eqn:gal_kernel}.

While we could use cross-correlations with spectroscopic samples to determine the amplitude of the bias evolution as well as $f(z)$, we instead use
a simpler and more robust method: cross-correlations
with gravitational lensing of the cosmic microwave background. Defining the effective bias $b^{\rm eff}$
\begin{equation}
b^{\rm eff} = \int \, dz \, b(z) \frac{dN}{dz} 
\label{eqn:beff}
\end{equation}
in the Limber approximation, we may write the angular
cross-correlation between the galaxies and CMB lensing
as \citep{Krolewski20}
\begin{multline}
    C_{\ell}^{\kappa g} = b^{\rm eff} \int d\chi \frac{W^{\kappa}(\chi)}{\chi^2} H(z) \left[f(z) \frac{dN}{dz}\right] 
    P(k \chi = \ell \, + \, 1/2) \\
    + \int d\chi \frac{W^{\kappa}(\chi) W^{\mu}(\chi)}{\chi^2} P(k \chi = \ell \, + \, 1/2)
\end{multline}

In Table~\ref{tab:galaxyselection_isw}, we give the best-fit $b^{\rm eff}$ for each sample.  The first set of errorbars are statistical error and the second set are error from uncertain $dN/dz$ (computed as the standard deviation of the best-fit bias from the 100 sampled $dN/dz$ as in Ref.~\cite{Krolewski20}).
The errors on the bias are at most 3\%, and are thus
negligible compared to the $\gtrsim 30\%$
errors on the ISW measurement (Table~\ref{tab:syst_check}).


While the dominant terms in the ISW cross-correlation (Eq.~\ref{eqn:isw}) require
$f(z) dN/dz$ rather than $dN/dz$, the lensing
magnification terms are sensitive to $dN/dz$ alone.
Here, we use the redshift distribution
of unWISE galaxies matched to optical sources in the deep
imaging of the 2 deg$^2$ COSMOS field. We use the multi-band photometric redshifts of Ref.~\cite{Laigle+15}, which have accuracy
$\Delta z/(1+z) = 0.007$ (referred to as ``cross-match redshifts,'' following \citep{Krolewski20}).
The COSMOS imaging is sufficiently deep
that nearly all of the unWISE galaxies
have counterparts with photometric redshifts.

\begin{figure}
    \centering
    \resizebox{1.0\columnwidth}{!}{\includegraphics{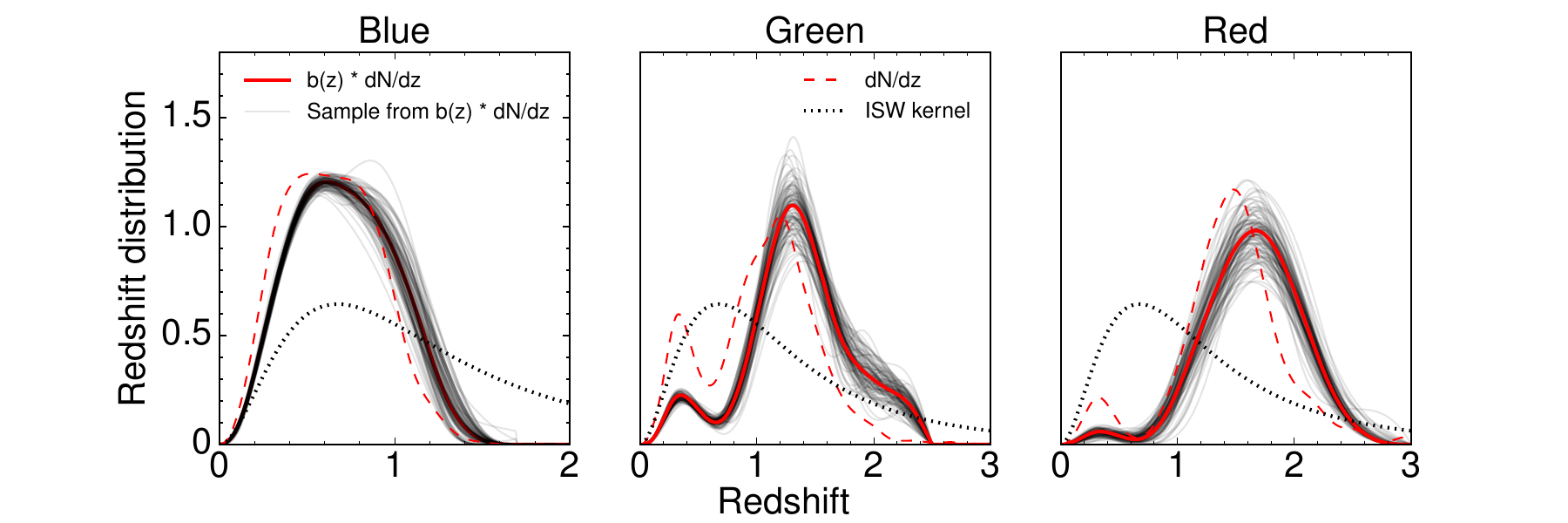}}
    \caption{Redshift distributions of the 3 unWISE samples.  The best measurement of the unWISE redshift
    distribution comes from cross-correlations
    with spectroscopic samples, which constrain
    the product of the bias evolution $b(z)$ and the redshift
    distribution $dN/dz$ (solid red lines).  The cross-correlation redshift
    measurement is noisy and realizations of the uncertainty
    in $b(z) dN/dz$ are shown as the gray lines.
    We also show the redshift distribution $dN/dz$
    as determined from cross-matching
    to multi-band photometric redshifts in the COSMOS field (red dashed line). $dN/dz$ enters in the ISW-magnification term
    (Eq.~\ref{eqn:isw_mag}). As this term has a subdominant
    contribution to the overall signal, the uncertainties
    from uncertain $dN/dz$ are small and therefore not shown.  The ISW kernel, $\frac{d[(1+z) D(z)]}{dz} \frac{dV}{dz}$, is shown as the black dotted line.
    }
    \label{fig:dndz}
\end{figure}

The cross-match redshifts are shifted to lower
redshift than the cross-correlation redshifts (Fig.~\ref{fig:dndz}).
We show in Ref.~\cite{Krolewski20} that reconciling
the cross-match and cross-correlation redshift
distributions requires a bias evolution
consistent with the observed evolution in the unWISE
number density. We further construct in Ref.~\cite{Krolewski21} a plausible Halo Occupation Distribution model that matches the COSMOS
$dN/dz$, the clustering $f(z) dN/dz$, $C_{\ell}^{\kappa g}$ and $C_{\ell}^{gg}$.
Therefore, we conclude that the COSMOS $dN/dz$
and clustering $f(z) dN/dz$ are consistent with each other.

Since the cross-correlation redshift distribution
 comes from noisy clustering measurements,
the redshift distribution $f(z) dN/dz$ has uncertainty as well.
We create samples of $f(z) dN/dz$ that are consistent
with the data and whose density is proportional to their probability of being the correct one given the data. These samples are obtained by generating
Gaussian random realizations with the correct noise covariance (one for each sample), adding the noise realization to the measured $f(z) dN/dz$, and
finally fitting a smooth B-spline with positivity
constraint and curvature penalty. The method is
described further in Ref.~\cite{Krolewski20},
and also used in Ref.~\cite{Krolewski21}.

The cross-match redshift distribution is also noisy,
with errors arising from uncertain photometric redshifts,
sample variance, cosmic variance, and potentially
variations in unWISE galaxy properties across the sky (since COSMOS is only 2 deg$^2$).
Due to the diverse sources of the noise, the uncertainties
in cross-match $dN/dz$ are less well characterized.
Moreover, as we show in Section~\ref{sec:ISW_measurement},
the uncertainty in $f(z) dN/dz$ is subdominant
to the statistical uncertainty on the ISW measurement.
Since the magnification terms
sensitive to cross-match $dN/dz$ are smaller
than the clustering terms sensitive to $f(z) dN/dz$,
the contributions from uncertainty in $dN/dz$
are negligible compared to the statistical uncertainty.

An additional potential source of systematic error comes from the scale mismatch between the $C_{\ell}^{Tg}$ measurement and the cross-correlation redshift measurement.
We measure the cross-correlation redshifts on fairly small scales (2.5 to 10 $h^{-1}$ Mpc in configuration space) so the bias will not be identical to the linear bias on large
scales appropriate for ISW.
It is therefore possible that the quasi-linear bias that we measure in the cross-correlation redshifts evolves differently with redshift as the linear bias.  We construct simple HODs for the WISE sample \citep[Appendix B in ][]{Krolewski20}, and we can gain some understanding into the nonlinear bias evolution of the WISE sample using $N$-body simulations populated with these HODs.  We find that the systematic shift and error from nonlinear bias evolution is smaller than the uncertainty from the measurement error in $dN/dz$. We optionally apply the ``nonlinear bias correction,'' derived from the HODs, as a correction to $f(z) \frac{dN}{dz}$ and find that it does not have a significant effect on the results.



\subsubsection{Comparison to previous WISE samples}
\label{sec:compare_to_prev_samples}

\begin{figure}
    \centering
    \resizebox{1.0\columnwidth}{!}{\includegraphics{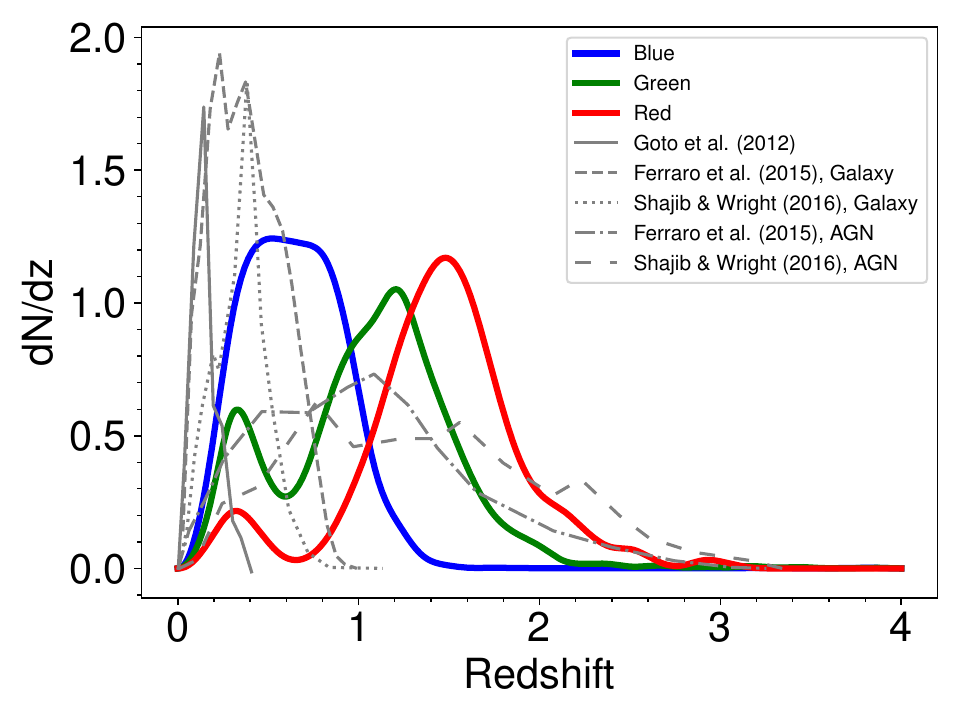}}
    \caption{Redshift distribution of the galaxy samples used in this work (thick lines) to previous WISE samples (thin gray lines).  The galaxy and AGN redshift distributions used by Ref.~\cite{Ferraro15}
    are from Ref.~\cite{Yan13} (galaxies) and Ref.~\cite{Geach13} (AGN),
    and these redshift distributions
    were also used in Ref.~\cite{Planck2015_ISW}.
    Other non-WISE galaxy samples often
    have much smaller sky coverage
    and are either at $z \lesssim 0.7$ or have a broad redshift kernel similar to the AGN samples.
    }
    \label{fig:dndz_compare}
\end{figure}

Many previous studies have used WISE
samples to measure the ISW effect
\cite{Goto12,Ferraro15,Shajib16,Planck2015_ISW}. These works have often selected two samples
of WISE tracers, a galaxy sample at $z \sim 0.5$ and an AGN sample with a broad kernel out to $z \sim 2$. In contrast, this work
provides more tomographic information on ISW
than these previous samples, as we split
the unWISE galaxies into three samples
at $z \sim 0.5$, 1.1, and 1.5. The lowest
redshift blue sample significantly overlaps
with these  previous samples, but even here
it pushes to somewhat higher redshift ($z_{\rm max} \sim 1.1$ vs.\ 0.8) than previous work (Fig.~\ref{fig:dndz_compare}).
Meanwhile, the green and red samples overlap
the AGN samples of Refs.~\cite{Ferraro15,Shajib16}, but are considerably narrower. 
Moreover, the green and red samples are considerably denser than the AGN samples, with densities 1868 and 144 deg$^{-2}$, respectively,
compared to 46 and 79 deg$^{-2}$ for the AGN samples from Refs.~\cite{Ferraro15} and \cite{Shajib16}, respectively.
Therefore our work is unique in that we provide an
ISW measurement in better localized bins at $z = 1.1$ and 1.5 over two thirds of the sky, with considerably higher galaxy density than previous work.
Many ISW measurements have been made with non-WISE samples as well, but these typically
probe smaller sky areas (e.g.\ tomographic quasar correlations in Ref.~\cite{Stolzner18}), are limited to $z \lesssim 0.7$ (e.g.\ 2MPZ \cite{Rassat07,Planck2015_ISW}, SDSS \cite{Planck2015_ISW,Stolzner18}, ATLAS \cite{Ansarinejad20}, WISE x SuperCOSMOS \cite{Stolzner18}),
or have very broad redshift distributions (NVSS \cite{Planck2015_ISW}).
Hence while our measurement is correlated
with previous work, it achieves tighter
constraints on the ISW amplitude in
an interesting redshift range at the transition
between matter and dark energy domination.

\subsection{Planck CMB data}
\label{sec:isw_planck}

We use the SMICA CMB temperature map from the Planck 2018 release as our fiducial temperature map in the ISW analysis.\footnote{Obtained from the Planck Legacy Archive, \url{http:/pla.esac.esa.int}}
We use the common confidence mask (combined confidence mask for the different temperature pipelines) as described in Section 4.2 of Ref.~\cite{Planck2018_compsep}.
SMICA produces a temperature map from a linear combination of the Planck input channels (30 to 857 GHz) with multipole-dependent
weights, up to $\ell \sim 4000$.
We repeat the measurement with the COMMANDER; NILC; SEVEM multi-frequency; SEVEM 70, 100, 143 and 217 GHz; and SMICA-noSZ maps to test the robustness of the result \citep{Planck2018_compsep}.
To measure the covariance, we use 300 FFP10 end-to-end simulations released as part of the Planck 2018 data release
\citep{Planck2018_2,Planck2018_3,Planck2018_compsep}.
These simulations include instrumental noise,
systematics, and foregrounds,
and were processed with an identical
pipeline to the data to produce component-separated maps.

\begin{figure}
    \centering
    \resizebox{1.0\columnwidth}{!}{\includegraphics{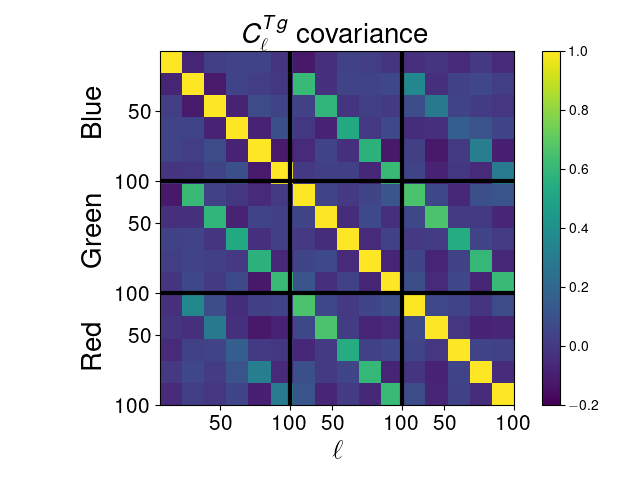}}
    \caption{Correlation matrix of $C_{\ell}^{Tg}$, as determined
    by repeating the cross-correlation measurement
    on 300 Planck full-sky mocks
    and unWISE galaxy data.
    The bins shown are the fiducial ones used for the analysis; blue, green and red share 5 bins at $20 < \ell < 100$, and blue also uses an additional bin at $7 < \ell < 20$ which is omitted for green and red
    due to concerns about systematics in the galaxy sample.
    }
    \label{fig:covariance}
\end{figure}

For the CMB lensing cross-correlations
used to measure $b^{\rm eff}$, we use the minimum-variance (MV) CMB
lensing maps obtained from temperature
and polarization \textsc{SMICA} maps
from the Planck 2018 release. We mask the
field with the mask provided
by the Planck team.
Our methodology closely follows
Ref.~\cite{Krolewski20} with two minor
differences\footnote{These are the same analysis choices used in Ref.~\cite{Krolewski21} for $C_{\ell}^{\kappa g}$.}: we use $20 < \ell < 1000$
rather than $\ell_{\rm min} = 100$ (as this conservative $\ell_{\rm min}$ is only necessary for the galaxy auto-correlation, which we do not use in this work);
and
we set $f(z) dN/dz$ to zero at $z > 1.5$ (2.5, 3)
for blue (green, red). Since the cross-match
$dN/dz$ is nearly zero beyond these limits,
any bumps in $f(z) dN/dz$ are likely to be noise.

\section{Angular power spectrum measurement}
\label{sec:isw_measurements}


In order to estimate the binned cross and auto power spectra, we use a pseudo-$C_\ell$ estimator \citep{Hivon02} based on the harmonic coefficients of the galaxy and temperature fields.
We follow the same procedure as in Ref.~\cite{Krolewski20}.
The measured pseudo-$C_\ell$ on the cut sky are calculated as 
\begin{equation}
\Tilde{C}_\ell^{XY} = \frac{1}{2\ell + 1}\sum_m X_{\ell m} Y^{\star}_{\ell m}
\label{eqn:pseudo_cl}
\end{equation}
where $X,Y\in\{g_1,g_2,g_3,T,\kappa\}$ are the observed fields on the cut sky.
Because of the mask, these differ from the true $C_\ell$ that are calculated from theory, but their expectation value is related through a mode-coupling matrix, $M_{\ell \ell'}$, such that 
\begin{equation}
\langle \Tilde{C}_\ell \rangle = \sum_{\ell'} M_{\ell \ell'}  C_{\ell'}
\label{eq:mode_coupling}
\end{equation}
The matrix $M_{\ell \ell'}$ is purely geometric and can be computed from the power spectrum of the mask itself. While Eq.~(\ref{eq:mode_coupling}) is not directly invertible for all $\ell$, the MASTER algorithm \citep{Hivon02} provides an efficient method to do so assuming that the power spectrum is piecewise constant in a number of discrete bins, $b$.
Defining a ``binned'' mode-coupling matrix, $\mathcal{M}_{b b'}$ \citep{Alonso18}, we can recover unbiased binned bandpowers 
\begin{equation}
   C_b = \sum_{b'} \mathcal{M}^{-1}_{b b'} \tilde{C}_{b'} \quad .
\label{eq:mode_coupling_binned}
\end{equation}
We use the implementation in the code \texttt{NaMaster}\footnote{\url{https://github.com/LSSTDESC/NaMaster}} \citep{Alonso18}.

We mask the galaxy map with the unWISE mask including bright stars and galaxies, and the CMB map with the ``common'' mask for CMB temperature, apodized
with a Gaussian smoothing kernel with FWHM 1 degree.
We also correct the unWISE density map by an ``area lost'' mask to account for the reduction in available area in each pixel due to point sources in Gaia (since we mask any source within 2.75'' of a star).
We test our pipeline on 100 noiseless Gaussian simulations (i.e.\ the CMB component is $C_{\ell}^{TT}$ from the late-time ISW only, and no galaxy shot-noise is added)  to ensure that we recover the correct power spectrum. 
The binned theory spectrum is the dot product of the unbinned theory spectrum and bandpower
window functions given by NaMaster. We check the bandpower window functions using Gaussian simulations
and find that they are correct to within the Poisson measurement error on the simulations
 (Appendix~\ref{sec:transfer}).


Since the azimuthal modes of the map are most affected by Galactic latitude-dependent foregrounds, we
remove the $m= 0$ mode from the sum in Eq.~\ref{eqn:pseudo_cl}.
This makes a very modest ($< 0.25\sigma$) impact on our results, and we validate
this procedure by omitting the $m=0$ modes in the Gaussian
simulation test. Despite the omission of the $m= 0$ mode, we find that the NaMaster bandpower window
function describes the effect of binning very well.

We omit large-scale modes where the auto-spectrum of the unWISE
galaxies deviates significantly from a theory model fit to smaller scales.
We find that the auto-spectrum contains significant spurious power below $\ell = 7$ for blue and $\ell = 20$ for green and red. Thus we use
bins at $7 < \ell < 100$ for blue and $20 < \ell < 100$
for green and red.
For all samples, we use bins with width $\Delta \ell = 16$ from $20 < \ell < 100$, and we add a 
bin at $7 < \ell < 20$ for blue.
This is a conservative choice for $\ell_{\rm min}$, as the systematics
 in the galaxy map are likely not correlated with potential residual
 systematics in CMB temperature.  Indeed we find that lowering $\ell_{\rm min}
$ to 5 for the green sample leads to negligible changes in $A_{\rm ISW}$.
However, since these modes are quite noisy, conservatively omitting
them from the analysis makes little difference in our results.
We omit modes at $\ell > 100$, where the ISW signal
is small, consistent with $\ell_{\rm max}$
commonly used in previous work \citep{Ferraro15,Shajib16}.

We use the 300 Planck simulations to determine the covariance of the ISW power spectra, including the cross-covariance
between the different galaxy samples (Fig.~\ref{fig:covariance}). We apply our pipeline to measure the cross-correlation between each simulation
and the unWISE maps, and then measure the covariance of these 300 power spectra.
We apply the Hartlap correction to the inverse covariance matrix \cite{hart+06} to account for noise in the mock-based covariance, although this correction is tiny due to the small size of the data vector (5 bins).
We find that the error bars from the Planck mocks are generally quite similar to the Gaussian error bars \citep{Knox95, Hivon02, Efstathiou:2003dj}.
As a further check, we find that the error bars from the Planck mocks are generally similar to the scatter of the individual $C_{\ell}$ within
each bin.

\section{Comparison to theory}
\label{sec:ISW_measurement}

\subsection{ISW measurement and comparison to $\Lambda$CDM}

\begin{figure}
    \centering
    \resizebox{1.0\columnwidth}{!}{\includegraphics{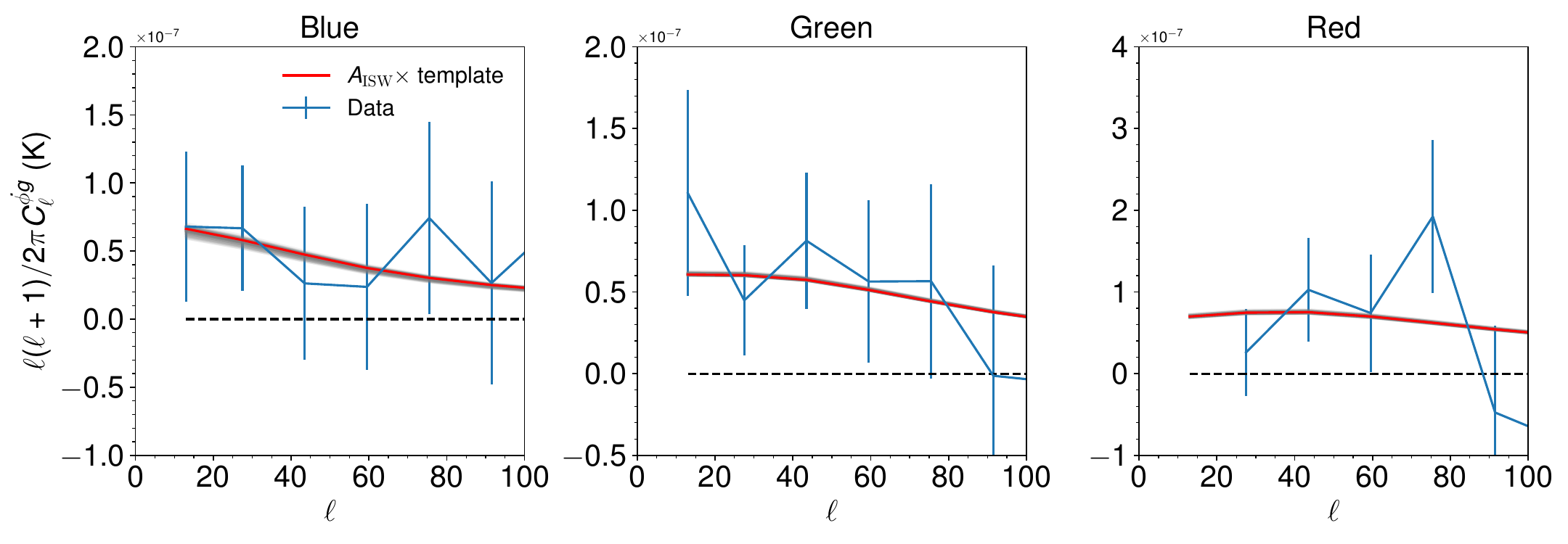}}
    \caption[Integrated Sachs-Wolfe effect from unWISE-Planck cross correlation]{ISW data (blue)
    and prediction in the $\Lambda$CDM Planck cosmology, scaled by $A_{\rm ISW}$ (red) for the three unWISE samples.
    Errors are from the mock-based covariance in Fig.~\ref{fig:covariance}.
    The gray range around the theory curve is the uncertainty
    propgated from errors in
    $b dN/dz$.
    }
    \label{fig:isw}
\end{figure}

In Fig.~\ref{fig:isw}, we show the measured ISW cross-correlation and the $\Lambda$CDM theory curve in the fiducial cosmology, multiplied by a scaling factor $A_{\rm ISW}$. Specifically, we bin the $C_{\ell}^{\dot{\Phi}-{\rm unWISE}}$
template identically to the data;
multiply theory and measurement
by $\ell_{\rm center} (\ell_{\rm center} + 1)/2\pi$ to work with data
that is roughly constant (where $\ell_{\rm center}$
is the center of each bin);
and multiply the template by the bandpower window function (Fig.~\ref{fig:transfer}).

Fig.~\ref{fig:isw} demonstrates
that our data is consistent with $\Lambda$CDM.
We find $A_{\rm ISW} = 0.73 \pm 0.34$ for blue ($2.1\sigma$), $1.00 \pm 0.39$ for green ($2.6\sigma$), and $1.14 \pm 0.52$ for red ($2.2\sigma$), for a combined significance 
of $3.2\sigma$ (combined $A_{\rm ISW} = 0.96 \pm 0.30$), again using the 300 FFP10 simulations
to calculate the correct covariances between the individual
ISW cross-correlations.
The $\Lambda$CDM model ($A_{\rm ISW}$ = 1) provides a good fit to the data,
with $\chi^2 = 10.4$ over 15 degrees of freedom.

\begin{figure}
    \centering
    \resizebox{1.0\columnwidth}{!}{\includegraphics{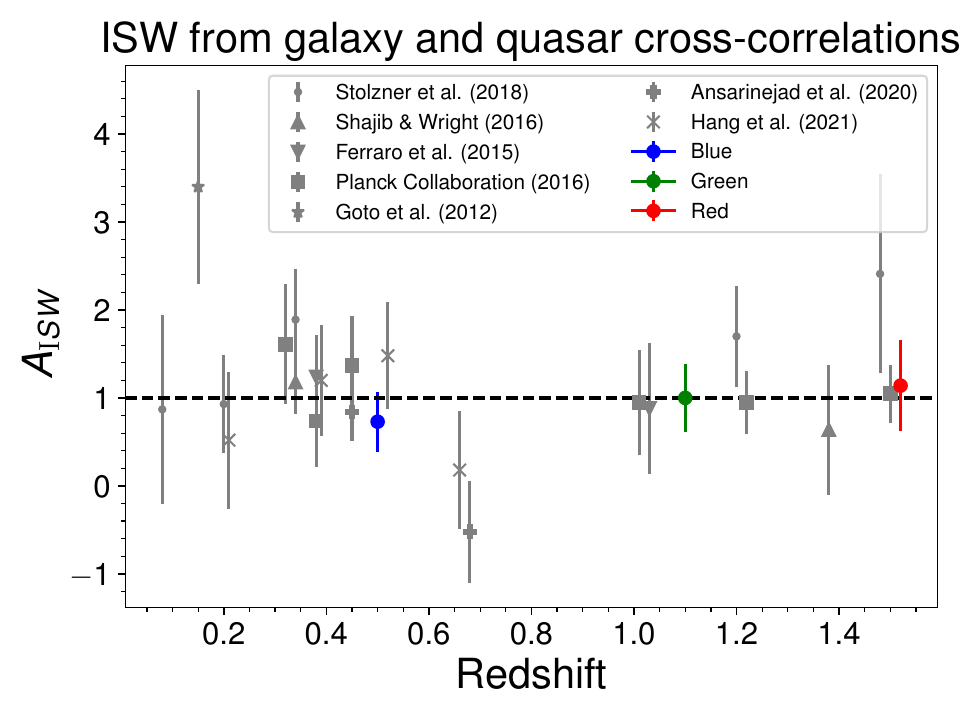}}
    \caption{Comparison of our results (colored points) to previous constraints on $A_{\rm ISW}$ from ISW cross-correlations (gray) from Refs.~\cite{Stolzner18,Shajib16,Ferraro15,Planck2015_ISW,Goto12,Ansarinejad20,Hang20}. We have taken all $A_{\rm ISW}$ measurements directly from the papers without attempting to homogenize the cosmologies used. For Ref.~\cite{Stolzner18},
    we have used the combined $A_{\rm ISW}$ for each survey
    rather than for individaul redshift bins, to present a secure detection. For Ref.~\cite{Planck2015_ISW}, we use
    the SEVEM results. For Ref.~\cite{Ansarinejad20},
    we present the ATLAS+SDSS results in two bins, $z = 0.35$ plus 0.55, and $z = 0.68$.  We note that many of these measurements use very similar or even identical datasets,
    and in some cases the redshift kernels are quite broad
    so a single mean redshift does not accurately represent the entire range covered. Moreover, even apart from duplicated
    datasets these points are highly covariant. We have only included ISW measurements from cross-correlations with galaxies and quasars, and do not show measurements from cross-correlations with super-clusters or voids \cite{Kovacs19,Kovacs21,Hang21}.  We omit these measurements as they have much larger errorbars and often less secure detections, though we note that they find hints of tension with $\Lambda_{\rm CDM}$.
    }
    \label{fig:all_ISW}
\end{figure}

The significance of detection for the blue sample
is somewhat lower than previous expections.
Ref.~\cite{Afshordi04} expects signal-to-noise
of $\sim 5$ for an $f_{\rm sky} = 0.6$ survey
with 100 million galaxies out to $z_{\rm max} = 1$, with bias of 1.5.  However, our $\ell_{\rm min}=7$ cut reduces the signal-to-noise by $\sim 20$\% (the $\ell_{\rm max} = 100$ cut makes $<2\%$ difference).  Furthermore, even with the
$\ell_{\rm min}$ cut, the autocorrelation
of the blue sample is higher than the $\Lambda$CDM expectation (presumably due to uncorrected systematics). Without this excess noise at low $\ell$, an ISW signal with $A_{\rm ISW} = 1$ would be detected at 3.7$\sigma$ (versus $1/0.34 = 2.9\sigma$ from Table~\ref{tab:syst_check}).
This is quite consistent with the 4$\sigma$
detection expected from Ref.~\cite{Afshordi04}.
In addition, our measurement is consistent with past ISW
cross-correlation measurements (Fig.~\ref{fig:all_ISW}).

\begin{table}[]
\centering
\begin{tabular}{c|ccc}
Analysis & Blue $A_{\rm ISW}$ & Green $A_{\rm ISW}$ &  Red $A_{\rm ISW}$ \\
\hline
\rowcolor{gray!20}
SMICA & $0.73 \pm 0.34$ & $1.00 \pm 0.39$ & $1.14 \pm 0.52$ \\
COMMANDER & $0.73 \pm 0.34$ & $0.99 \pm 0.39$ & $1.11 \pm 0.52$ \\
SEVEM & $0.76 \pm 0.34$ & $1.01 \pm 0.39$ & $1.15 \pm 0.52$ \\
NILC & $0.74 \pm 0.34$ & $1.00 \pm 0.39$ & $1.13 \pm 0.52$ \\
SMICA-NoSZ & $0.83 \pm 0.34$ & $1.07 \pm 0.39$ & $1.17 \pm 0.52$ \\
SEVEM 70 GHz & $0.66 \pm 0.34$ & $0.91 \pm 0.39$ & $1.12 \pm 0.52$ \\
SEVEM 100 GHz & $0.75 \pm 0.34$ & $0.97 \pm 0.39$ & $1.10 \pm 0.52$ \\
SEVEM 143 GHz & $0.76 \pm 0.34$ & $1.00 \pm 0.39$ & $1.14 \pm 0.52$ \\
SEVEM 217 GHz & $0.87 \pm 0.34$ & $1.04 \pm 0.39$ & $1.13 \pm 0.52$ \\
$m = 0$ included& $0.69 \pm 0.34$ & $0.88 \pm 0.39$ & $1.03 \pm 0.52$ \\
Weighted galaxy field & $0.73 \pm 0.34$ & $0.98 \pm 0.39$ & $1.12 \pm 0.52$ \\
\end{tabular}
\caption[Systematics tests in ISW cross-correlation]{$A_{\rm ISW}$ for the three unWISE
samples for a variety of different systematics tests: changing the method used
to construct the CMB temperature maps; 
including the $m = 0$ mode;
and applying weights to the galaxy field.
The default analysis is shown on the top line with 
gray highlighting.
}
\label{tab:syst_check}
\end{table}

In Table~\ref{tab:syst_check}, we perform a variety of systematics checks. We replace the default SMICA map with several
other CMB temperature maps \citep{Planck2018_compsep}, and do not find that using any of these maps lead to a significant
shift in $A_{\rm ISW}$.  We also show that the results are not significantly
affected by 
applying
weights to the galaxy field or including the $m= 0$ mode.

The uncertainty in $b dN/dz$ creates a systematic uncertainty
in the theoretical prediction.
We compute $A_{\rm ISW}$ using theoretical templates generated
from each of the 100 samples of $b dN/dz$, with the bias taken 
from the corresponding fit of $C_{\ell}^{\kappa g}$ to that sample's
redshift distribution.  We find uncertainties on $A_{\rm ISW}$
of 0.025, 0.0106, and 0.0180 for blue, green, and red, respectively.
These systematic uncertainties are much smaller than the statistical
errors on $A_{\rm ISW}$.

As Table~\ref{tab:galaxyselection_isw}
shows, the statistical uncertainties
on $b^{\rm eff}$ are small (of order $1-2\%$)
and thus contribute a negligible
theoretical uncertainty compared
to the large statistical errors on $A_{\rm ISW}$.
Furthermore, the systematic uncertainties
from nonlinear biasing are similar to the statistical
uncertainties: if we restrict to $\ell < 300$ (instead
of the fiducial $\ell < 1000$) we find $b^{\rm eff} = 1.54$, 2.27 and 3.16 for the three samples, versus $b^{\rm eff} = 1.50$, 2.23 and 3.19 if we use all data with $\ell < 1000$.
In Ref.~\cite{Krolewski20}, we also study
other systematic
errors on the redshift distribution and
find them to be generally
smaller than the errors from uncertain $dN/dz$.
Therefore, we expect them to make a negligible
contribution the ISW errorbars.

\subsection{Comparison to dynamical dark energy}

``Freezing'' and ``tracking''
type behavior in the dark energy equation of state, $w$ (corresponding to $w = -1$ at late times and $w = 0$ at early times, respectively), may be generic features of single scalar field theories \citep{Raveri17,Bull21}.
Since these models cause dark energy to act like matter at early times, increasing its energy density, they modify the ISW signal. Our measurement is particularly sensitive to these models if the transition from freezing to tracking
occurs at $z \sim 1$ where the galaxy kernels peak.

As a specific example, we consider the Mocker model
of Ref.~\cite{Linder06,Linder06b}, which is a phenomenological description imitating the behavior
of quintessence models. The Mocker model is defined by
\begin{equation}
    \frac{dw}{d \log{a}} = Cw(1+w)
\end{equation}
yielding the following equation of state, with free
parameters $C$ and $w_0$
\begin{equation}
    w(a) = -1 + \left[1 - \frac{w_0}{1 + w_0} a^C\right]^{-1}
\end{equation}
$w_0$ sets the $z = 0$ equation of state, well constrained
to be close to $-1$ by supernovae and BAO. $C$ controls the transition to a matter-like
equation of state, with larger values of $C$ yielding
a transition at lower redshift.
The cosmological constant is recovered by $w_0 = -1$.
Observational constraints on this model have been considered in Ref.~\cite{Xia09,Xia09b,Bull21}.

\begin{figure}
    \centering
    \resizebox{1.0\columnwidth}{!}{\includegraphics{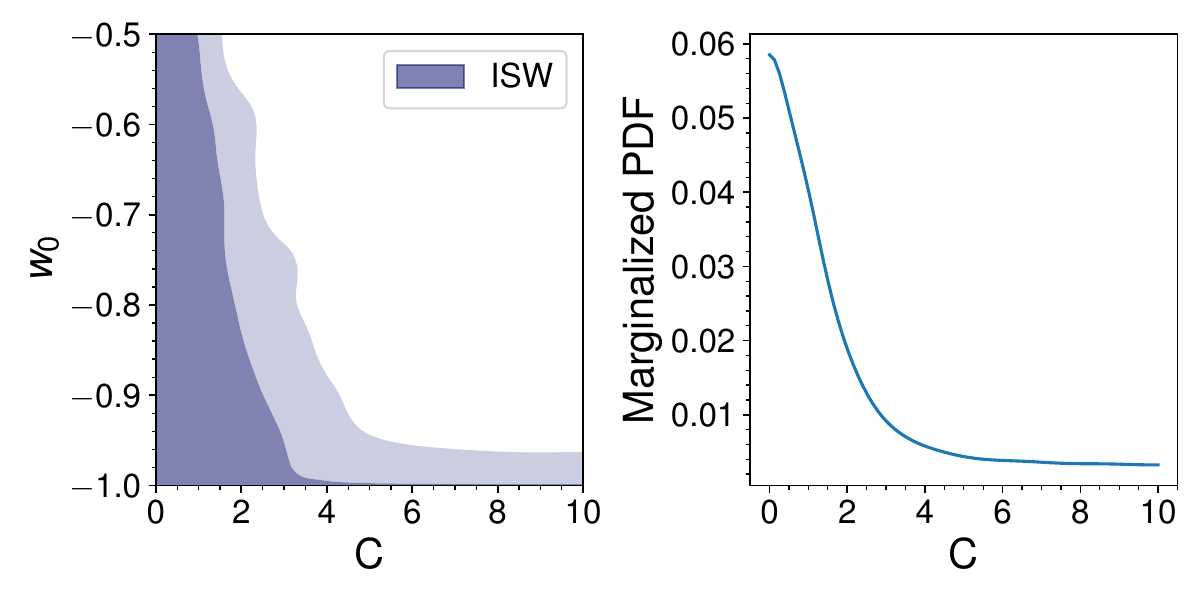}}
    \caption{Left panel: one and two $\sigma$ constraints on mocker parameters $C$ and $w_0$ from unWISE ISW alone. The $w = -1$ axis corresponds to $\Lambda$CDM.
    Right panel: marginalized constraints on $C$ from unWISE ISW.  Marginalized
    constraints on $w_0$ are much weaker
    than evidence from geometric probes
    that the Universe is dark energy dominated at low redshift. The ISW measurement
    provides more information on the evolution of the dark energy component through the parameter $C$.
    }
    \label{fig:mocker1}
\end{figure}

We consider constraints on the Mocker parameters
from the ISW measurement in Fig.~\ref{fig:mocker1}.
We impose a flat prior on $C$ between 0 and $\infty$ and on $w_0$ between $-1$ and 0.
We fix the other cosmological parameters to a flat cosmology
with parameters identical to the ones used in the previous section: $\Omega_m h^2 = 0.1417$, $\theta_{\rm MC} = 0.0104$, $\Omega_b h^2 = 0.0224$, $n_s = 0.9665$, $\sigma_8 = 0.8102$, and one massive neutrino at 0.06 eV.
We choose this parameterization as it is best
constrained by the Planck primary CMB observations on smaller scales than the ones considered here. We use \textsc{CAMBSources} to compute $C_{\ell}^{Tg}$ in the Mocker model, using the \textsc{DarkEnergyPPF}
class \cite{Fang08} to allow for an arbitrary $w(a)$.

The shape of the ISW posterior is determined by the fact that all models at $w_0 = -1$ are equivalent, regardless
of $C$; hence the posterior expands as $w_0$ approaches $-1$ and models with different $C$ become increasingly
similar. The long non-zero tail in $C$ corresponds
to models in which the transition from dark energy to matter occurs at very low redshift, 
where the ISW measurement is less sensitive
due to the drop in the galaxy redshift distribution at $z < 0.5$.  ISW measurements
from lower-redshift samples, or other measurements
of the expansion history, can rule
out these models with large $C$.  For instance,
the combination of unWISE ISW and Type Ia supernovae from Pantheon \cite{Scolnic17}
can improve constraints on the Mocker models from Pantheon alone. If we also include BAO measurements,
the constraints on Mocker models from the expansion history further improve
the constraints.  

\begin{figure}
    \centering
    \resizebox{1.0\columnwidth}{!}{\includegraphics{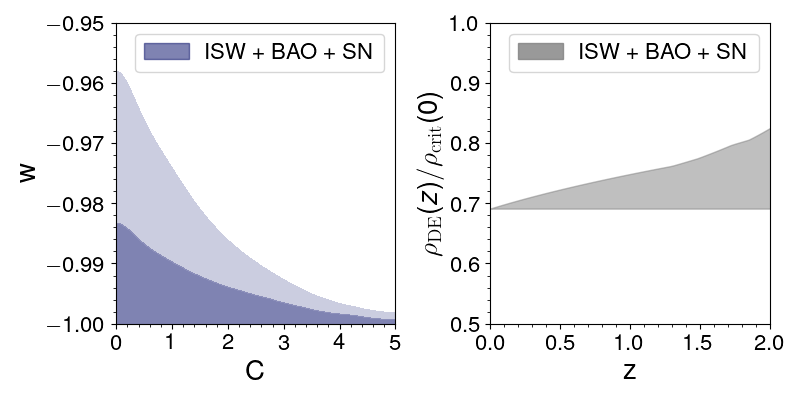}}
    \caption{Left panel: constraints on mocker parameters $C$ and $w_0$ from ISW and from type Ia supernovae plus BAO.  The $x$ axis corresponds to $\Lambda$CDM.
    Right panel: corresponding 95\% upper limit on the dark energy density as a function of redshift, in the Mocker model.
    }
    \label{fig:mocker2}
\end{figure}

In Fig.~\ref{fig:mocker2}, we show constraints from ISW and expansion
history measurements. We use the Pantheon compilation of Type Ia supernovae
\cite{Scolnic17}
and BAO from SDSS I-III (including Ly$\alpha$ forest auto-correlation and quasar cross-correlation) and 6dF \cite{Beutler12,Ross15,Alam21}.
Note that BAO and supernovae are uncalibrated standard
rulers and candles; i.e.\ BAO constrains the ratio of the 
comoving distance to the sound horizon (and thus does not
depend on knowing the absolute value of the sound horizon).
The ISW measurement improves the constraint from supernovae alone, and when adding BAO, the constraints further tighten significantly.
Our constraints improve upon those presented
in \cite{Xia09,Xia09b}; we find similar
constraints on $C$ (marginalized 95\% upper limit of $3.99$) and considerably improve
the constraint on $w_0$ (marginalized 95\% upper limit of $-0.97$). These represent
the tightest constraint on the Mocker model to date, and constrain the dark energy density
to be within $\sim 10\%$ of a cosmological constant (95\% upper limit) at $z = 2$ in these models.

\section{Conclusions}

Using Planck CMB temperature maps
and unWISE galaxy maps over 60\% of the sky,
we detect the integrated Sachs-Wolfe
correlation between galaxy density
and CMB temperature at 3.2$\sigma$.
We use three galaxy samples out to $z \sim 2$,
and find $A_{\rm ISW} = 0.73 \pm 0.34$ in the $z \sim 0.5$ ``blue'' sample, $A_{\rm ISW} = 1.00 \pm 0.39$ in the $z \sim 1.0$ ``green'' sample,
and $A_{\rm ISW} = 1.14 \pm 0.52$ in the $z \sim 1.5$ ``red'' sample.
We find that this detection is robust to a number
of changes in the analysis choices,
suggesting it is not significantly affected by systematics.
We also use this measurement
to constrain a toy phenomenological model
of freezing quintessence, the Mocker model.
We find
that the ISW measurement improves
constraints on the Mocker model compared to type Ia supernovae,
and adding BAO constraints we can obtain the tightest constraints.
We provide updated constraints
on the Mocker model with the latest BAO and supernovae datasets,
constraining the dark energy density to within 10\% at $z < 2$.

We apply a number of tests to ensure
the robustness of the ISW measurement.
First, we include the lensing magnification-$\dot{\Phi}$ term in our theoretical
modelling. This term is substantial for the green
and red samples, 15-20\% of the $g\dot{\Phi}$ term. Second, we remove $m = 0$ modes (in a Galactic coordinate system) as these are most
likely to be contaminated by systematics in the galaxy and ISW maps.  We test and validate
our pipeline on Gaussian mocks, and 
confirm the accuracy of the bandpower
window function required to generate the binned theory
power spectrum.
Third, we conduct a variety of systematics
tests, changing the CMB temperature map used,
the minimum $\ell$ used, and applying weights to the galaxy field, as summarized in Table~\ref{tab:syst_check}. Finally we measure
the covariance of the ISW signal by cross-correlating the galaxy maps with 300
end-to-end Planck simulations with fully realistic noise.

This measurement represents a direct detection of the effects
of dark energy, consistent with the best-fit
$\Lambda$CDM cosmology from Planck
with no statistically significant
evidence for evolution in the dark energy density.
Our results are quite consistent
with previous ISW cross-correlation results \citep{Giannantonio08,Planck2015_ISW,Stolzner18} and inconsistent with the higher ISW amplitude
reported using stacking around superclusters
or supervoids \citep{Grannett08,Kovacs19}.  
Compared to previous results,we measure the ISW signal
in narrower tomographic bins at $z \sim 1$ and 1.5 ($\Delta z \sim 0.5$ versus $\Delta z \sim 1$ from previous quasar and radio galaxy samples probing these redshifts), providing constraints that more sensitively probe the beginning of the dark-energy dominated epoch.

Overall, these results
support the consensus flat $\Lambda$CDM cosmology and improve constraints
on the dark energy density at $z \sim 1-2$ by 30-40\%. While the ISW measurement is not as statistically significant as distance measurements
supporting dark energy (i.e.\ type Ia supernovae and BAO),
it complements them by constraining the time evolution of dark energy,
and offering a purely gravitational
rather than expansion-based constraint.

\section*{Acknowledgments}
We thank Martin White and Eddie Schlafly for many useful discussions on the unWISE and Planck data.
We thank Niayesh Afshordi, Martin White and Will Percival for useful comments that have improved this manuscript.
A.K.~is supported by the AMTD Foundation.
S.F.~is supported by the Physics Division of Lawrence Berkeley National Laboratory.
We acknowledge the use of \texttt{NaMaster} \cite{Alonso18}, \texttt{Cobaya} \cite{CobayaSoftware, Cobaya}, \texttt{GetDist} \cite{lewis2019getdist}, \texttt{CAMB} \cite{Lewis:1999bs} and thank their authors for making these products public. This research used resources of the National Energy Research Scientific Computing Center (NERSC), a U.S.\ Department of Energy Office of Science User Facility operated under Contract No.\ DE-AC02-05CH11231.
This research was enabled in part by software provided by Compute Ontario (\url{https://www.computeontario.ca}) and Compute Canada (\url{http://www.computecanada.ca}).
This work made extensive use of the NASA Astrophysics Data System and of the {\tt astro-ph} preprint archive at {\tt arXiv.org}.

\appendix

\section{Validating the bandpower window functions}
\label{sec:transfer}

\begin{figure}
    \centering
    \resizebox{1.0\columnwidth}{!}{\includegraphics{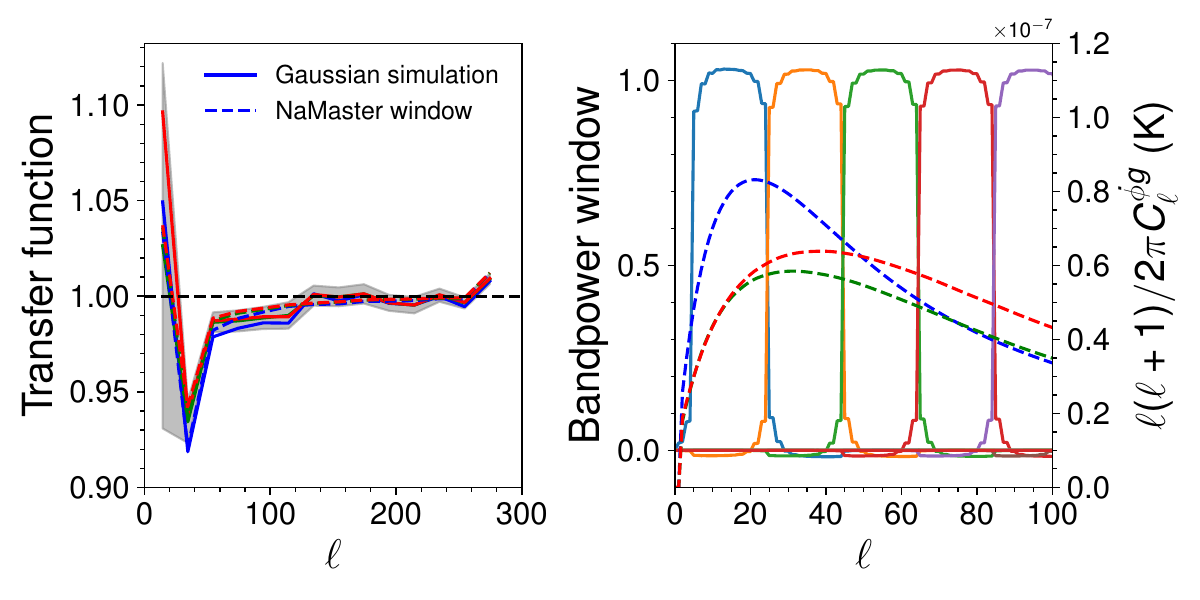}}
    \caption{\textit{Left:} Mask deconvolution transfer function for $C_{\ell}^{Tg}$ measurement, i.e. the ratio between
    the mean of the unbinned $C_{\ell}^{Tg}$, and the output binned power spectrum after masking,
    pseudo-$C_{\ell}$ estimation,
    and mask deconvolution.
    The transfer function was computed
    by averaging over 100 Gaussian simulations.
    $C_{\ell}^{gg}$ in the simulation
    is a smoothed version of
    the measured $C_{\ell}^{gg}$,
    including an uptick at low $\ell$
    above a theory curve fit to $C_{\ell}^{gg}$ at $\ell > 100$, presumably due to uncorrected systematics.
    The transfer function
    from the Gaussian simulation
    is compared to the product
    of the bandpower window
    matrix and the unbinned power spectrum (dashed line). The NaMaster window
    is perfectly consistent with the Gaussian simulations within the error on the simulations (gray band).
    \textit{Right:} The bandpower window functions (left axis) and the unbinned
    theory spectra (right axis).
    The product of these two quantities, divided
    by the mean of the unbinned spectrum in each bin,
    gives the dashed line
    in the left panel.
    }
    \label{fig:transfer}
\end{figure}

Due to the effects of mode-coupling,
the binned theory power spectrum is not identical
to the mean of the theory power spectrum
across each bin.
Instead, the binned power spectrum in bin $b$ is the product
of the bandpower window matrix $W$ and the unbinned power spectrum $C_{\ell}$
\begin{equation}
 C_{b} = \sum_{\ell} W_{b\ell} C_{\ell}
\end{equation}
The bandpower window matrix
is the product of the mode-coupling matrix $M_{\ell \ell'}$
and the binned mode-coupling
matrix $\mathcal{M}_{b b'}$
in Equations.~\ref{eq:mode_coupling} and~\ref{eq:mode_coupling_binned}
\begin{equation}
W_{b\ell} = \sum_{b'} \mathcal{M}^{-1}_{bb'} \sum_{\ell' \in b'} M_{\ell \ell'}
\end{equation}
The right panel of Fig.~\ref{fig:transfer} shows the bandpower
window matrix for the first five bins,
and the unbinned theory spectra on the right axis.
The left panel of Fig.~\ref{fig:transfer} measures
this effect empirically, by running the  measurement
pipeline on 100 Gaussian simulations and plotting the ratio
between the binned bandpowers (averaged over simulations)
and the mean of $C \textbf{}_{\ell}$ within each bin.
The measurement in simulations is fully consistent
with the expectation from the bandpower window
matrix; the apparent large discrepancy in the lowest $\ell$
bin is not statistically significant and entirely due to
cosmic variance in the bin.
The dip at $\ell = 35$ is expected
because the theory power spectra peak around $\ell = 35$;
thus the tails of the bandpower window function both 
decrease the binned power spectrum. At higher $\ell$, 
the two tails partially cancel, but because the spectrum
is a declining function of $\ell$, the binned power spectrum
is slightly below the mean of the unbinned power spectrum.

\bibliographystyle{JHEP}
\bibliography{main}

\end{document}